%% file: xinxin.tex
\documentclass[12pt]{article}

\input preamble.tex
\title{Computer-Aided Diagnosis of Low Grade Endometrial Stromal Sarcoma (LGESS)}

\author{Xinxin Yang\footnotemark[1]\ \ \ 
Mark Stamp\footnotemark[1]\,\,\footnotemark[2]}

\begin{document}

\symbolfootnotetext[1]{Department of Computer Science, San Jose State University}
\symbolfootnotetext[2]{mark.stamp$@$sjsu.edu}

\maketitle

\abstract
Low grade endometrial stromal sarcoma (LGESS) is rare form of cancer, 
accounting for about~0.2\%\ of all uterine cancer cases. 
Approximately~75\%\ of LGESS 
patients are initially misdiagnosed with leiomyoma, which is a type of benign 
tumor, also known as fibroids. 
In this research, uterine tissue biopsy images of potential LGESS patients 
are preprocessed using segmentation and staining normalization algorithms. 
A variety of classic machine learning and leading deep learning models 
are then applied to classify tissue images as either benign or cancerous. 
For the classic techniques considered, the highest classification accuracy 
we attain is about~0.85, while our best deep learning model achieves an 
accuracy of approximately~0.87.
These results indicate that properly trained learning algorithms
can play a useful role in the diagnosis of LGESS.

\section{Introduction}

Cancer is the second leading cause of death in the United States, 
accounting for~21.6\% of all deaths in a~2017 survey conducted by the 
Center for Disease Control (CDC)~\cite{aaaa}. The tremendous medical costs 
of cancer treatments and the harm the disease brings to patients and their families 
makes cancer an important area of medical research.

Low grade endometrial stromal sarcoma (LGESS) is a tumor comprised of 
endometrial stromal cells. It is rare, accounting for approximately~0.2\% of 
uterine cancers~\cite{a,b}. Most patients with LGESS have a good prognosis, 
with a 5-year survival rate of about~80\% after surgical removal of the tumor. 
However, LGESS has a relatively high recurrence rates. at about~60\%, 
and the disease-related death rate is estimated to be between~15\% and~25\%~\cite{c,d}. 

When diagnosing LGESS, it is difficult to differentiate LGESS from benign leiomyoma, 
which is also known as fibroids.  Only~10\% of LGESS patients are correctly diagnosed, 
whereas~75\%\ are misdiagnosed with preoperative leiomyoma~\cite{e}. Many cases 
even remain misdiagnosed postoperative~\cite{f}. More accurate and automatic image 
analysis methods would be useful to diagnose LGESS, assess treatment efficacy, 
and lower cancer-related costs to the healthcare system.

Researchers in artificial intelligence (AI) rarely 
possess the medical knowledge required for accurate modeling a disease
such as LGESS. And AI algorithms that are less than~100\%\ accurate
make such techniques more suited as an aid in cancer risk assessment, 
rather than definitive diagnostic tools~\cite{ff}. Learning models can reduce 
the workload of healthcare professionals by automating tedious tasks, 
such as tumor segmentation. Moreover, AI algorithms may be more capable than 
humans at analyzing smaller, more subtle structures in patient images~\cite{bb}. 
Also, computers can analyze larger feature sets in a shorter amount of time, potentially
allowing for a more nuanced analysis of image structure that is not easily perceived 
by a human viewer. These capabilities have been showcased in a wide variety of use 
cases, including tumor segmentation, determination of tumor malignancy, and prediction 
of survivability in afflicted patients~\cite{bbb}.

In this research, we apply machine learning and deep learning methods to classify soft 
tissue images of potential LGESS patients. Our goal is to provide tools for improved 
diagnostic accuracy of LGESS tumors.

The remainder of this paper is organized as follows: Section~\ref{chap:background} 
contains relevant background material, including a
review of research literature related to cancer image analysis,
and a brief introduction to the machine learning and deep learning algorithms
that we employ in our LGESS experiments. Specifically, we consider
classic machine learning techniques of multilayer perceptron (MLP),
random forest, XGBoost, support vector machines (SVM), 
principal component analysis (PCA) in combination with SVM.
In the realm of deep learning algorithms, we consider 
convolutional neural networks (CNN), residual networks (ResNet),
AlexNet, and DenseNet.

Section~\ref{chap:dataset} gives an 
overview of the cancer image dataset used in this research, and 
we discuss the preprocessing strategies that we employ. 
This data preprocessing workflow includes region of interest (ROI) segmentation, 
image patch extraction, and stain normalization. In Section~\ref{chap:experiments}, 
we provide details on our machine learning and deep learning experiments,
and we analyze the results. 
Section~\ref{chap:conclusion} concludes the paper, 
and we provide suggestions for possible future work 
to help diagnose LGESS based on image analysis.

\section{Background}\label{chap:background}

In this section, we discuss relevant background topics.
First, we discuss relevant related work, then we provide
a brief overview of each of the classic machine learning
and deep learning algorithms that we employ for the analysis
of LGESS images.

\subsection{Related work}

Machine learning has found widespread use in cancer classification and 
diagnosis~\cite{bbb, bba}. Although machine and deep learning research 
has been conducted on many different cancers, we are aware of
no existing studies pertaining to LGESS. This may be due to the rarity 
of LGESS, as compared to other cancers that are typically studied in
this type of research.

Mesrabadi and Faez~\cite{g} apply artificial neural networks (ANN), 
AlexNet, and support vector machines (SVM) to classify prostate cancer. 
In their experiments, AlexNet yielded a classification accuracy of~86.3\%, 
compared to~81.1\%\ for SVM and~79.3\%\ from ANNs. 
Kharya et al.~\cite{h} note that generic ANNs are the most widely used prediction technique 
in medical forecasting, but the structure of such models is difficult to understand. 
Kharya et al. also consider the relative advantages and disadvantages of decision trees, 
na\"{i}ve Bayes, ANNs and SVMs for breast cancer detection.

Ashhar et al.~\cite{hh} consider the efficacy of deep learning algorithms for early 
detection of lung cancer. Lung cancer is almost always 
diagnosed at advanced stages---early and accurate screening would 
be beneficial to patient outcomes.  Ashhar et al. apply five 
state-of-the-art CNN-based architectures to a dataset of 
computer tomography (CT) lung cancer images. They find 
that GoogleNet gives the best results, with an accuracy, specificity, sensitivity, 
and AUC of~94.53\%, 99.06\%, 65.67\%, and~86.84\%, respectively.

Vijayarajeswari et al.~\cite{hha} apply Hough transforms to mammogram images 
to detect features that are potentially indicative of breast cancer. The images are
then classified using SVM. They attain a~94\%\ accuracy with this strategy, 
far surpassing the classification accuracy of SVM on unmodified images. 
The Wisconsin diagnostic breast cancer (WDBC) dataset was used as the
source of their images~\cite{hhb}. 

In~\cite{hhh}, Ghoneim et al. conduct research on cervical cancer detection and 
classification---cervical cancer is one of the leading causes of cancer death 
among women. They extract relevant features from cervical cancer images 
using CNNs,  then classify the images using extreme learning machines (ELM), 
multi-layer perceptrons (MLP) and autoencoder (AE) based classifiers. 
The best performance in these experiments came from a CNN-ELM-based 
system, which attained an accuracy of~99.5\%\ accuracy on 
a binary classification problem and~91.2\%\ accuracy on 
a multiclass problem.

Chaturvedi et al.~\cite{hhhh} propose a classification method for skin cancer 
that achieves better evaluation results than previous studies and exceeds
human dermatologists. Their implementation of the MobileNet model achieved
an overall accuracy of~83.1\%\ for multiclass experiments involving seven classes.

Bharat et al.~\cite{i} apply traditional machine learning classifiers,
including $k$-nearest neighbor ($k$-NN), na\"{i}ve Bayes, 
classification and regression trees (CART), and SVMs to 
the problem of predicting and diagnosing breast cancer. They conclude that 
these machine learning algorithms behave quite differently depending on 
the application. Specifically, they find that $k$-NN has the best 
diagnostic results, while na\"{i}ve Bayes and logistic regression 
perform well when applied to specific breast cancer diagnosis problems.  
These results generally agree with the research of Rana et al.~\cite{j}.

According to Maglogiannis et al.~\cite{k},
SVM is the best technique for predicting recurrence or non-recurrence 
of breast cancer.  In their research, SVM attains an accuracy of~96.91\%, 
specificity of~97.67\%, and sensitivity of~97.84\%, 
outperforming the Bayesian and ANN classifiers it was compared against. 

The results of these studies paint a promising picture for the predictive abilities 
of machine learning and deep learning in the cancer domain. In this research,
we consider deep learning and machine learning for the diagnosis of LGESS
based on image analysis.

\subsection{Learning Algorithms}

Based on our literature review, we will consider both classic machine learning models 
and various cutting-edge deep learning techniques. In this section, we provide 
a brief description of each learning model that we employ in our classification
experiments.

\subsubsection{Multilayer Perceptron}

Multilayer perceptrons (MLP) are a type of artificial neural network (ANN). 
An MLP includes an input layer, an output layer, with one or more 
hidden layers in between. MLPs can form the basis for so-called 
deep neural networks (DNN), where multiple hidden layers are used~\cite{nn}.

\subsubsection{Random Forest}

A random forest combines multiple decision trees, where each component
decision tree is trained on a subset of the data and features~\cite{nnn}. 
Every decision tree is itself a classifier, and hence if~$N$ decision trees 
comprise a random forest, we will have~$N$ classification results. A random forest 
typically uses a simple voting scheme among the component decision
treed to determine the classification.

\subsubsection{XGBoost}

Extreme gradient boosting (XGBoost)~\cite{nnnn}
has attracted considerable attention because of its 
efficiency and high prediction accuracy. 
XGBoost is based on gradient boosted decision trees (GBDT)
combined with various techniques designed to increase the 
efficiency, without compromising the accuracy.

\subsubsection{SVM}

Support vector machines (SVM) date to the early~1960s,
but only became practical in the mid-1990s~\cite{nnnnn}. 
Prior to the recent re-emergence of deep learning, SVM was regarded 
as one of the most successful machine learning algorithms, and it
is still competitive in many problem domains.

In a binary classification problem,
an SVM attempts to construct a separating hyperplane between the
classes, while maximizing the ``margin,'' i.e., the minimum distance between the 
hyperplane and the training samples. The strength of SVMs derive largely from
the so-called kernel trick, which enables us to work in a higher 
dimensional space---where there is a better chance of the data being
linearly separable---without paying a significant performance penalty.

\subsubsection{PCA}

Principal component analysis (PCA) is one of the most widely used 
dimensionality reduction algorithms~\cite{nnnnnn}. The PCA technique is
based on eigenvalue analysis, with the underlying assumption
being that linear combinations of features with larger variances 
are more relevant for classification than those with smaller variance.

\subsubsection{CNN}

\begin{figure}[!htb]
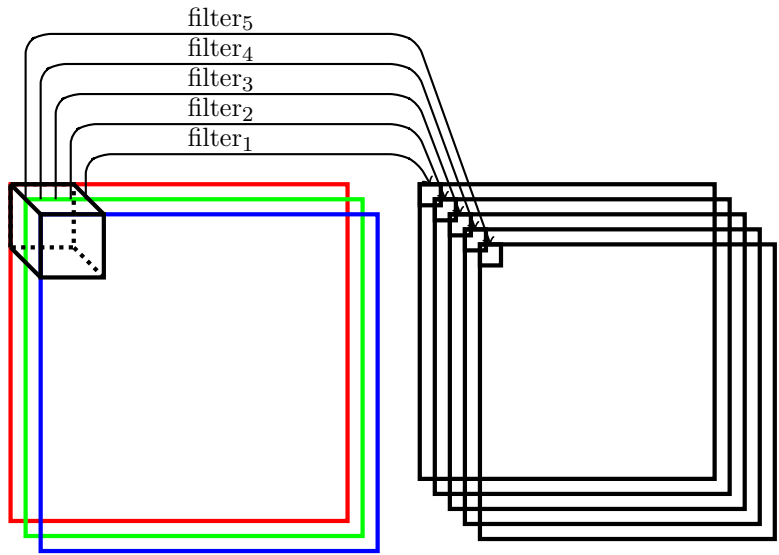

    \centering
        \input figures/conv3b.tex
    \caption{Convolution}\label{fig:CNN}
\end{figure}

As the name suggests, convolutional neural networks (CNN)
include convolutional layers, which consist of filters that
enable the efficient analysis of local structure~\cite{nnnnnnn}. 
Figure~\ref{fig:CNN} illustrates a case where five convolutions
are applied to three-dimensional data.

By applying multiple convolutional layers, progressively higher levels
of abstraction are obtained. 
CNNs were designed for image analysis, but the technique
can perform well in any task where local structure dominates. 

\subsubsection{AlexNet}

\begin{figure}[!htb]
        \centering
        \includegraphics[width=120mm]{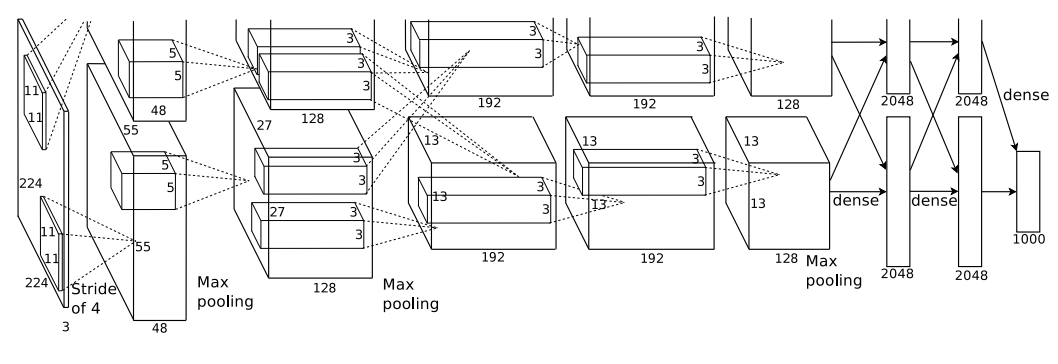}
        \caption{AlexNet architecture}
        \label{fig:AlexNet}
    \end{figure}

AlexNet is a specific CNN architecture. 
The AlexNet architecture is illustrated in Figure~\ref{fig:AlexNet},
as given in~\cite{p}.

From Figure~\ref{fig:AlexNet}, we see that AlexNet has eight 
layers, not including pooling layers, 
with the first five being convolutional layers, while the 
remainder are fully connected. There are~650,000 neurons 
and some~60 million parameters in AlexNet. See~\cite{p} for 
additional details on the AlexNet architecture.

\subsubsection{DenseNet}

\begin{figure}[!htb]
        \centering
        \includegraphics[width=150mm]{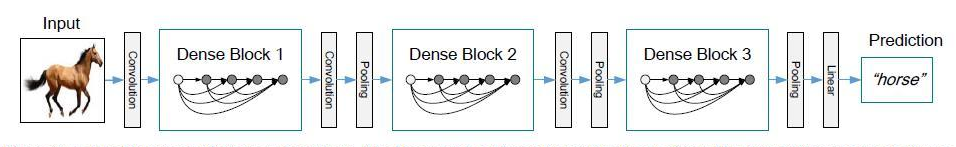}
        \caption{DenseNet architecture}
        \label{fig:DenseNet}
    \end{figure}

The dense convolutional network (DenseNet) architecture
was proposed in 2017~\cite{q}. 
DenseNet connects each layer in a feed-forward manner. 
In contrast to a typical convolutional neural network, in DenseNet,
the input of each layer includes the output of all previous layers.
Figure~\ref{fig:DenseNet} illustrates the DenseNet architecture,
as given in~\cite{q}.

\subsubsection{ResNet}

\begin{figure}[!htb]
        \centering
        \includegraphics[width=120mm]{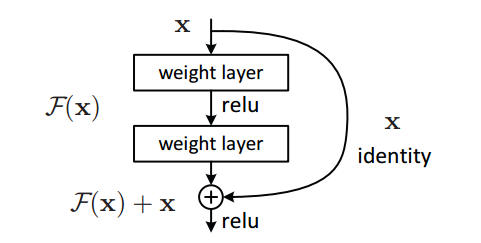}
        \caption{ResNet sub-network}
        \label{fig:ResNet}
    \end{figure}

Residual networks (ResNet) were first proposed in~2015
when a ResNet architecture won first place in the classification 
task for the ImageNet competition. ResNet
can be viewed as a collection of sub-networks that 
can be stacked to form a deep network. The use of these sub-networks,
or ``shortcut connections,'' enables efficient training of very deep networks~\cite{r}. 
Figure~\ref{fig:ResNet} illustrates a sub-network structure of ResNet.
The identity mapping in Figure~\ref{fig:ResNet} represents a
shortcut connection.

\section{Dataset and Preprocessing}\label{chap:dataset}

The dataset used in our experiments was obtained from 
the Cancer Imaging Archive, an organization offering data for cancer 
research~\cite{l}. Our specific dataset includes~888 uterine tissue biopsy 
images taken from~250 potential LGESS patients, formatted as SVS files. 
A separate annotations file contains the clinical diagnosis of each tissue image.

\begin{figure}[!htb]
    \centering
    \includegraphics[width= 120mm]{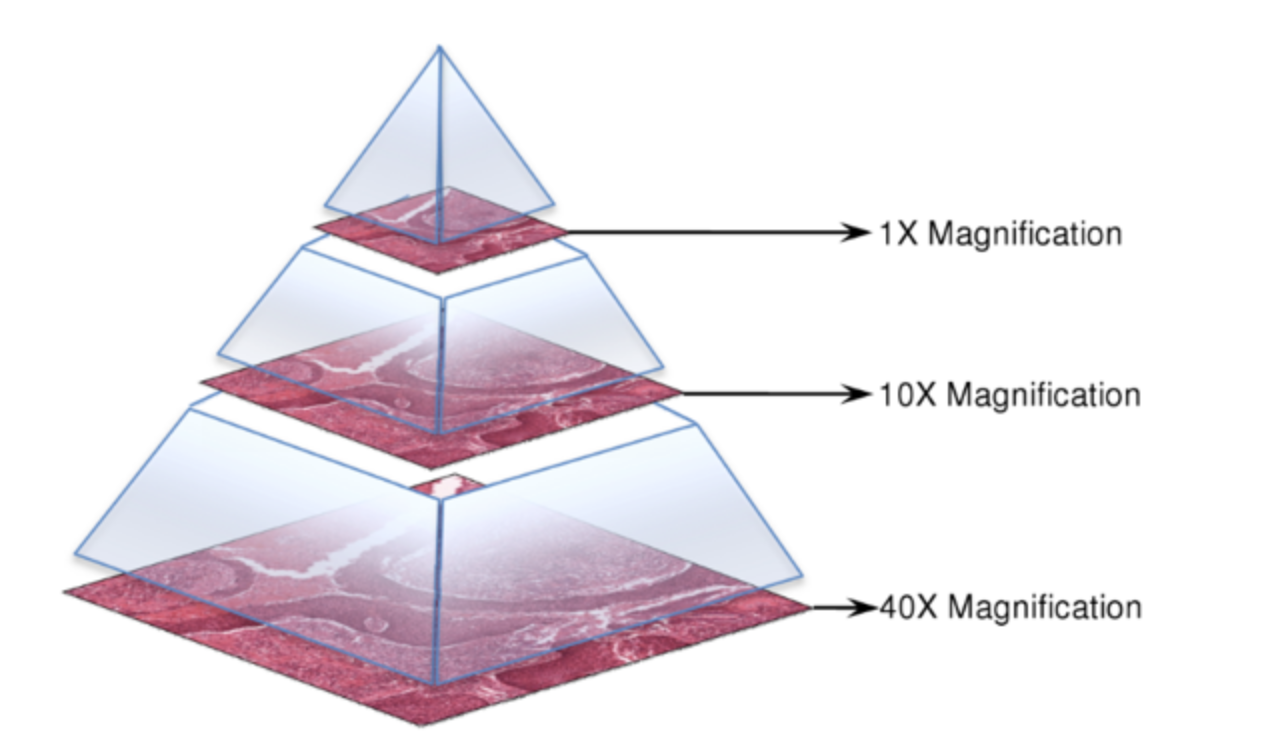}
    \caption{Whole slide image structure}
    \label{fig: WSI_struture}
\end{figure}

SVS is a file format for whole slide images (WSI). In a WSI, a microscope slide is 
scanned to create a single high-resolution digital file~\cite{m}. Most WSIs have 
a resolution of $100,000\times100,000$ pixels. WSIs are usually stored in a pyramid 
structure, where each level of the pyramid holds different downsampled versions of the 
original image~\cite{o}. The more ``downsampled'' an image, the lower the 
resolution and the less magnified it appears to be. Figure~\ref{fig: WSI_struture} 
illustrates the pyramid structure of a WSI image. 
In this research, we use the OpenSlide library~\cite{openslide}, 
which provides methods to read and access WSI images stored 
in a variety of file formats, including SVS.

WSIs offer clear visualization of tumor characteristics, including tissue infiltration, 
lymph node metastasis, and degree of differentiation. Such images are 
helpful for the diagnosis, prognosis, grading, and staging of tumors~\cite{n}.

The WSIs contained in the Cancer Imaging Archive dataset must undergo a 
color standardization stage before our classification algorithms can be applied,
since different production processes and scanning machines cause color variations. 
These color differences can cause problems for algorithms that are not robust to 
such variations, even if the differences are imperceptible to the human eye. 

The way we deal with these color difference is by color normalization,
which is also referred to as stain normalization. This involves normalizing all 
pictures to the color distribution of the a template picture~\cite{eee}. 

Before applying color normalization to our WSIs, the regions of interest (ROI) 
of each image need to be identified. This is done via standard Gaussian filtering 
and contour extraction techniques. The processing steps are explained in more detail below.

\begin{enumerate}
\item {\bf Segmentation of the target region from the image.}
\begin{enumerate}
\item Use the OpenSlide library to read the level~2 image of 
the WSI. This second level of the WSI pyramid still has good resolution, 
but contains less data and is thus easier and more efficient to process. 
We save the level~2 image in portable network graphics (PNG) format.
\item Transform the image from the RGB color space to grayscale.
\item Apply a Gaussian filter to normalize the image. 
We must make sure the filter threshold preserves the image contours.
\item Calculate the area of each contour and remove all contours that have an 
area below a specified threshold.
\item Obtain the final image mask based on the contours acquired from the previous steps. 
This image mask comprises the tissue regions. 
\end{enumerate}

The Figure~\ref{fig: contour_mask} below shows examples of extracted 
contours and the resulting image masks. In each row, the first image is the 
original image in the dataset, the second image is the mask we obtain from 
the original image, and the last image is the original image with the
contours marked.

\begin{figure}[!htb]
    \centering
    \includegraphics[width=120mm]{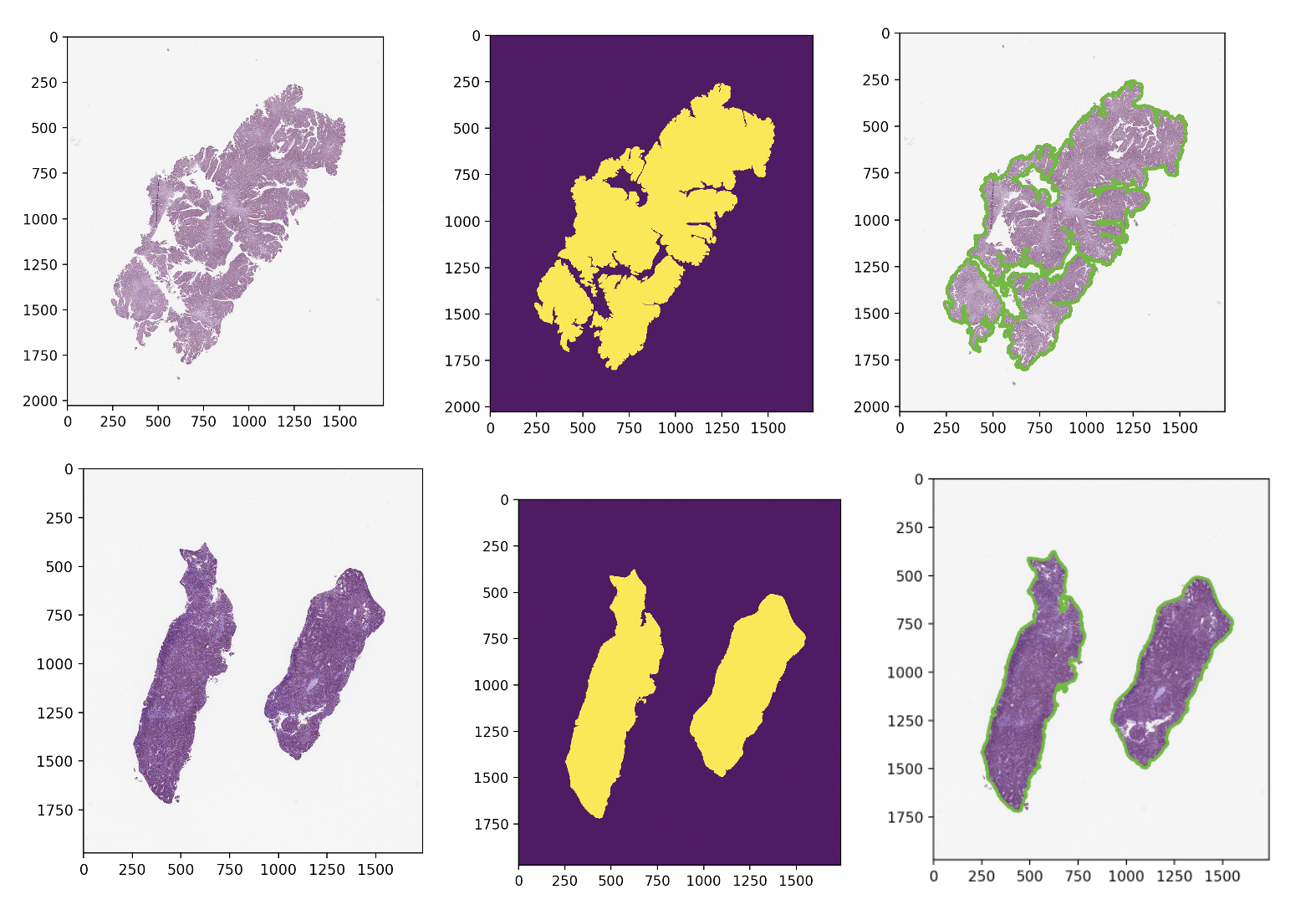}
    \caption{Contour and mask}
    \label{fig: contour_mask}
\end{figure}

\item {\bf Extract patches from an image's target region.}

We apply the previously-generated mask to the image, 
then walk through the image and cut patches of a predefined length and width. 
All patches extracted must have a certain area occupied by the masked image. 
Patches below this area threshold consist mostly of empty space and are therefore 
discarded. Examples of extracted patches are given in Figure~\ref{fig:patches}.

\begin{figure}[!htb]
    \centering
    \includegraphics[width=80mm]{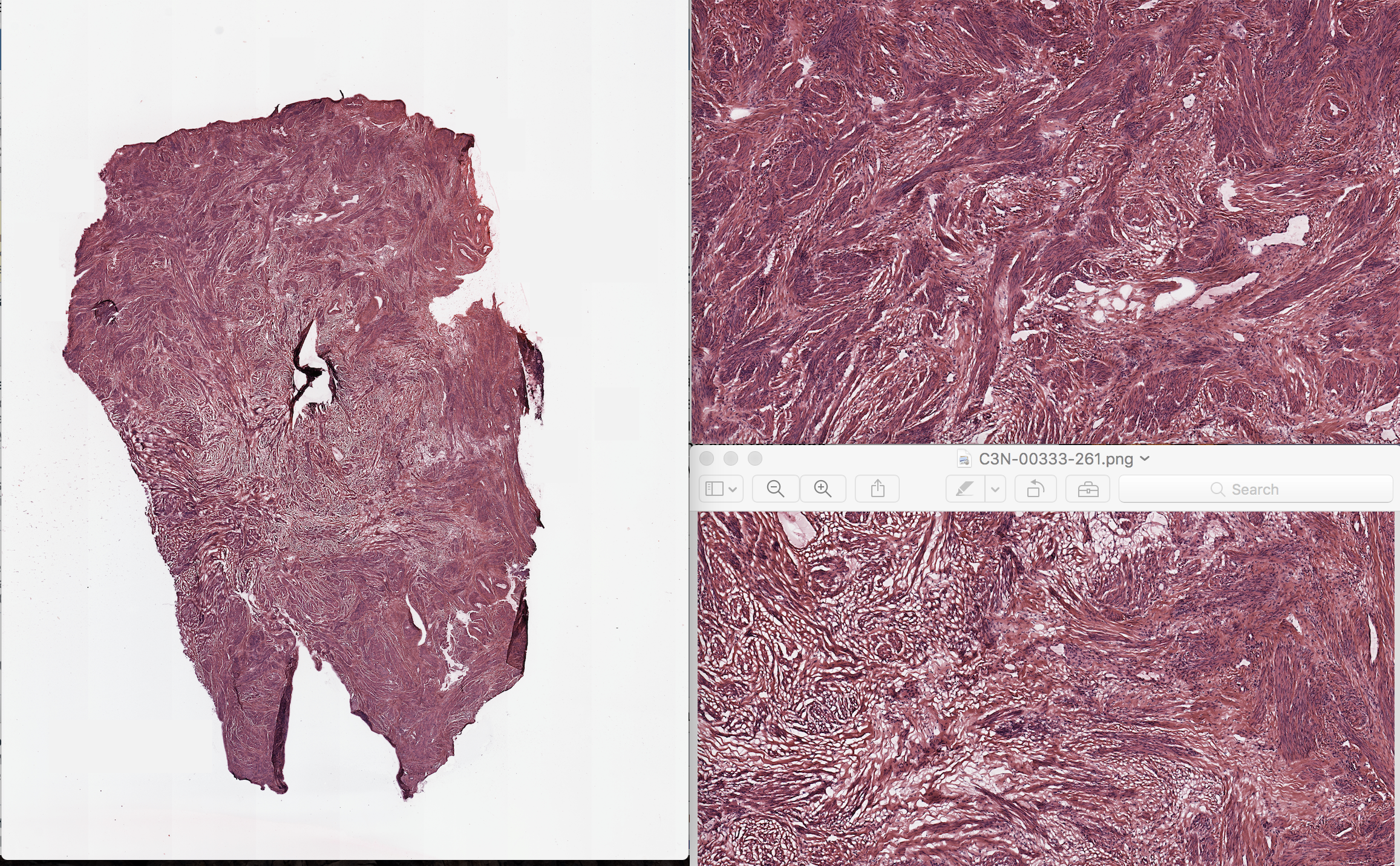}
    \caption{Patches cut from one image}
    \label{fig:patches} 
\end{figure}

\item {\bf Staining normalization.}

We apply Vahadane's staining normalization 
to achieve color normalization~\cite{eee}. The steps 
involved in this process are as follows:
\begin{enumerate}
    \item Optical density calculation.
    \item Unsupervised staining density estimation.
    \item Color normalization.
    \item Normalized pixel intensity calculation.
\end{enumerate}
Examples of images before and after color normalization are given 
in Figure~\ref{fig:color normalize for one}. 

\begin{figure}[!htb]
    \centering
    \includegraphics[width=115mm]{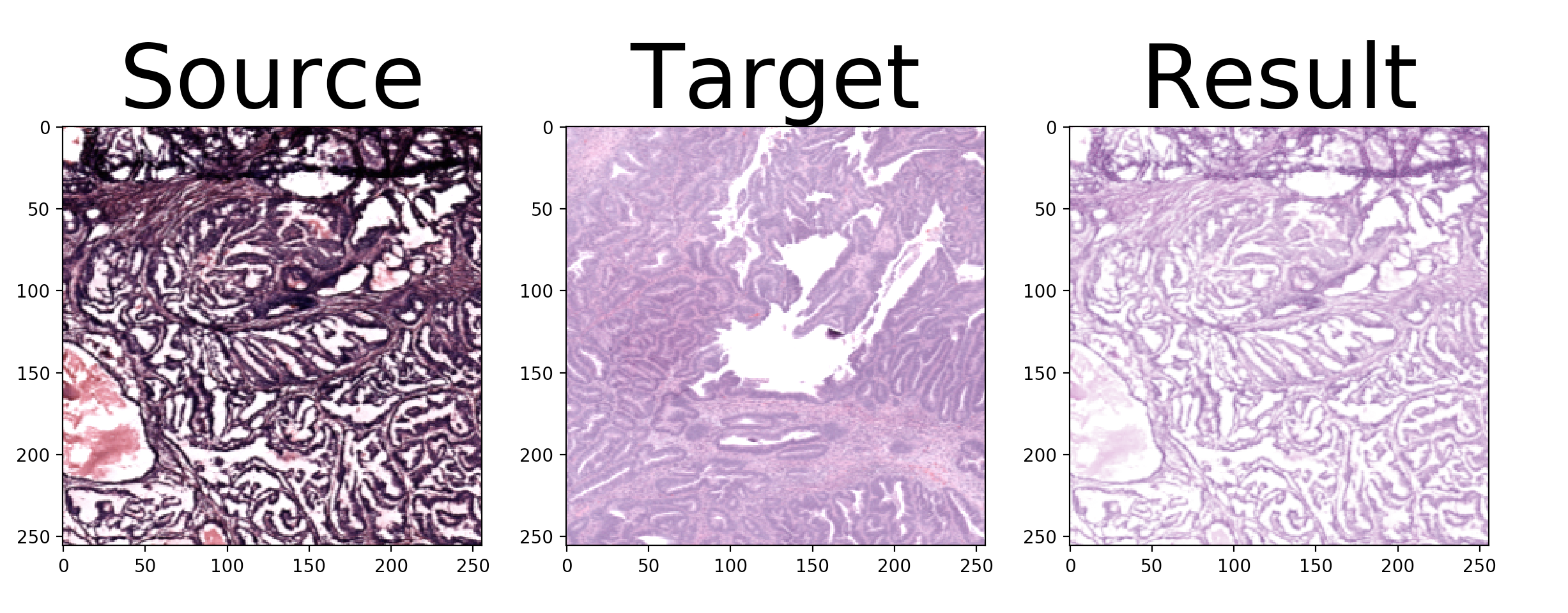}
    \caption{Color normalization of an image}
    \label{fig:color normalize for one} 
\end{figure}

\end{enumerate}

Figure~\ref{fig:color normalize for one} shows the color of an original image, 
labeled ``Source,'' while the image labeled ``Target'' indicates the color we would like 
our image to become. The image labeled ``Result'' shows the original source image 
after color normalization. Note that the color of the ``Result'' image is simialr
to that of the ``Target,'' as desired.

\begin{figure}[!htb]
    \centering
    \begin{tabular}{rc}\midrule\midrule
    \raisebox{0.45in}{Before:} & \includegraphics[width=115mm]{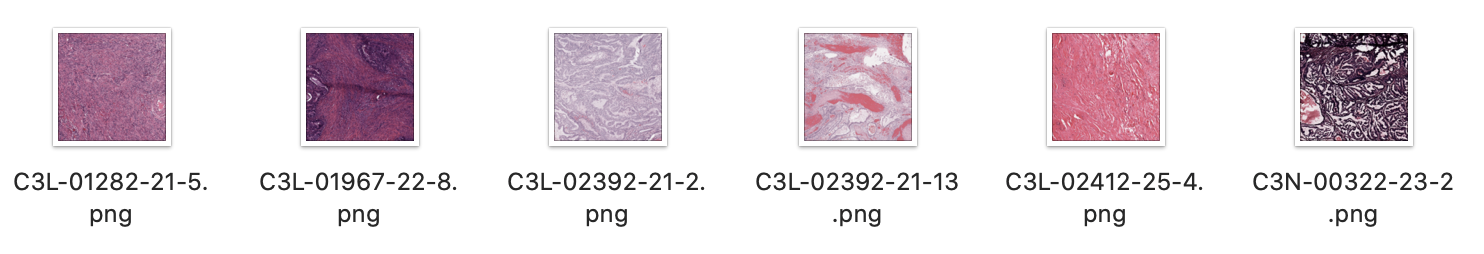} \\
    \raisebox{0.45in}{After:} & \includegraphics[width=115mm]{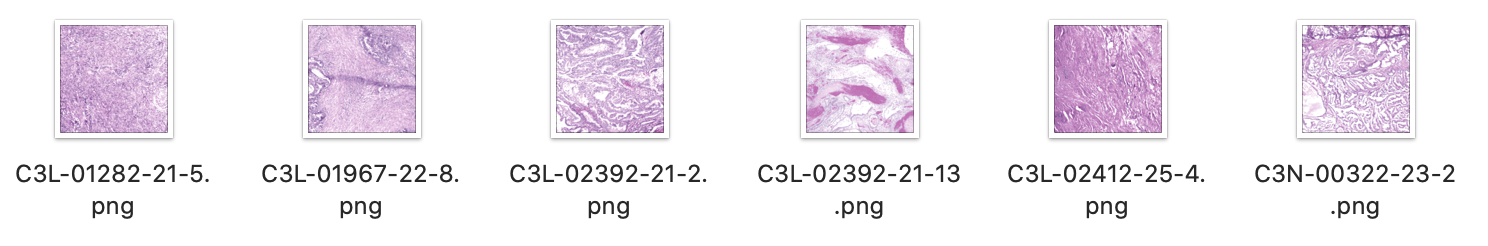} \\
    	\midrule\midrule
    \\
    \raisebox{0.45in}{Before:} & \includegraphics[width=115mm]{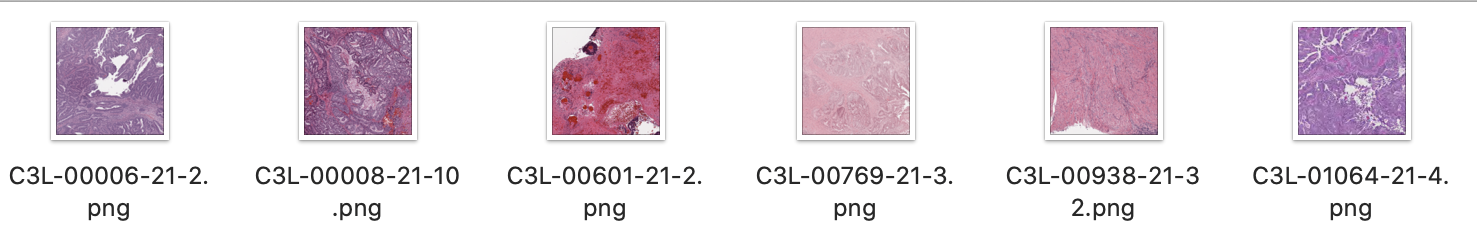} \\
    \raisebox{0.45in}{After:} & \includegraphics[width=115mm]{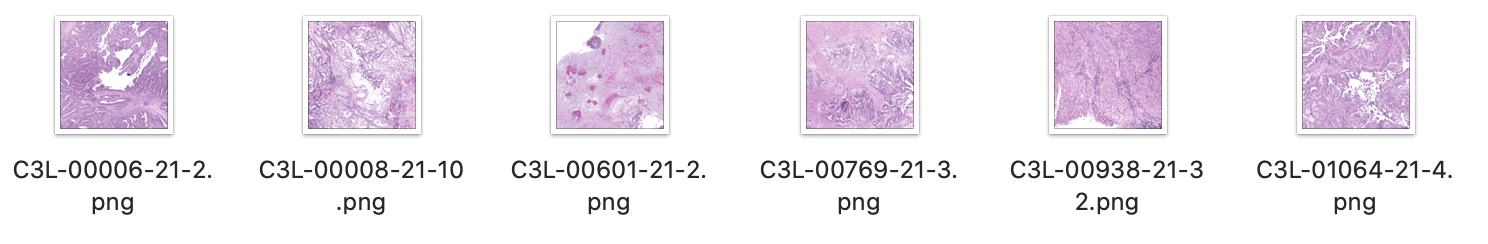} \\
    \midrule\midrule
    \end{tabular}
    \caption{Color normalization examples}
    \label{fig:color normalize comparison}
\end{figure}

Figure~\ref{fig:color normalize comparison} 
shows~12 image examples from our dataset before and after color normalization. 
Note that in each case,
the ``before'' image is directly above the corresponding ``after'' image.

Before normalization, we observe that the colors in the image can vary greatly. 
After our stain normalize process, the colors are clearly much more uniform.

\section{Experiments and Results}\label{chap:experiments}

After preprocessing the images and cutting them into patches, 
our dataset includes~4205 tumor images and~1459 benign images.
We reserve~20\% of the data as test data, while the remaining~80\% data is 
used as training data, with the training and test data randomly selected. We also used 5\-fold cross validation to avoid overfitting problems.
For each experiment, this gives us~4529 samples in the training set 
(3363 tumor images and~1166 benign images) and~1133 samples in 
the test set (841 tumor images and~292 benign images).

The metrics are used to evaluate the performance of our machine learning classifiers
are precision, recall, F1 score, and the accuracy. Precision is the proportion of true positives 
among those predicted as positive, while recall measures how well the positive samples 
are predicted. In cancer detection, it is imperative that models have a high recall. 
The~F1 score is the harmonic of the precision and recall, while accuracy 
is self-explanatory. 

For the basic techniques considered in the next section, we also
provide ROC analysis.
A receiver operating characteristic (ROC) curve is a plot of the true positive
rate versus the false positive rate as the threshold varies through all possible values.
The area under the ROC curve (AUC) can be 
interpreted as the probability that a randomly selected positive 
instance scores higher than a randomly selected negative instance.
It follows that the AUC provides a threshold-independent means of comparing
classifiers. An AUC of~1.0 implies that there is a threshold for which ideal separation is
attained, while an AUC of~0.5 means that the binary classifier is no better than a coin flip.
If an AUC value of~$0\leq x < 0.5$ is obtained, by simply switching the sense of
the classifier, we have an AUC of~$1-x > 0.5$, and hence the AUC can be no worse 
than~0.5.

\subsection{Basic Techniques}

In this section, we will discuss our experiment result on ``basic'' 
or standard machine learning techniques. Here we include as basic 
MLP, random forest, XGBoost, and SVM models, as well as a model
that combines PCA for dimensionality reduction with SVM.
Here, we discuss the specific architecture for each of these 
models and report the test accuracy. At the end of this section,
we provide a more detailed comparison of these basic models.

\subsubsection {Multilayer Perceptron}

In this experiment, we imported \texttt{MLPClassifier} 
from the \texttt{Scikit-Learn} library. We use a fully connected neural network 
with three hidden layers--the first, second, and third hidden layers have~600, 800, 
and~300 neuron, respectively. We use rectified linear units (RELU) 
as the activation function for each neuron an we train the model
for~600 epochs. In this case, we obtain a test accuracy of~0.5702.

\subsubsection{Random Forest}

We build our random forest model 
using \texttt{RandomForestClassifier} from the 
module \texttt{sklearn.ensemble}. 
We have experimented with the number of trees and find that~100 gives
the best accuracy.  Maximum depth of the tree was set to None, then the nodes were expanded until all leaves are pure or until all leaves contain less than 2. All other parameters of the model were set to default value.
In this case, we obtain a test accuracy of~0.7873.

\subsubsection{XGBoost}

Our XGBoost model is trained using the \texttt{xgboost} package
in Python. We obtain the best results witn \texttt{max\_depth} 
set to~6, and the objective function \texttt{binary:logistic}. 
We obtain a test accuracy of~0.8147.

\subsubsection{Support Vector Machine}

We use the SVM model from the \texttt{Scikit-Learn}. We found no significant difference
between any kernel functions tested, and hence for the sake of efficiency,
we select the linear kernel.
We set all other parameters to default except enable the probability estimates.
In our SVM experiment, we achieve a test accuracy of~0.8455.

\subsubsection{PCA with SVM}

In this experiment, we imported the PCA model from the \texttt{sklearn.decomposition} 
module. We combine the two models with the \texttt{make\_pipeline} method from 
the \texttt{sklearn.pipeline} module. We find that we obtain the best results 
with~300 components from PCA and with the rbf kernel in the SVM. 
In this model, we again obtain an test accuracy of~0.8535.

\subsubsection{Comparison of Basic Techniques}

\begin{figure}[!htb]
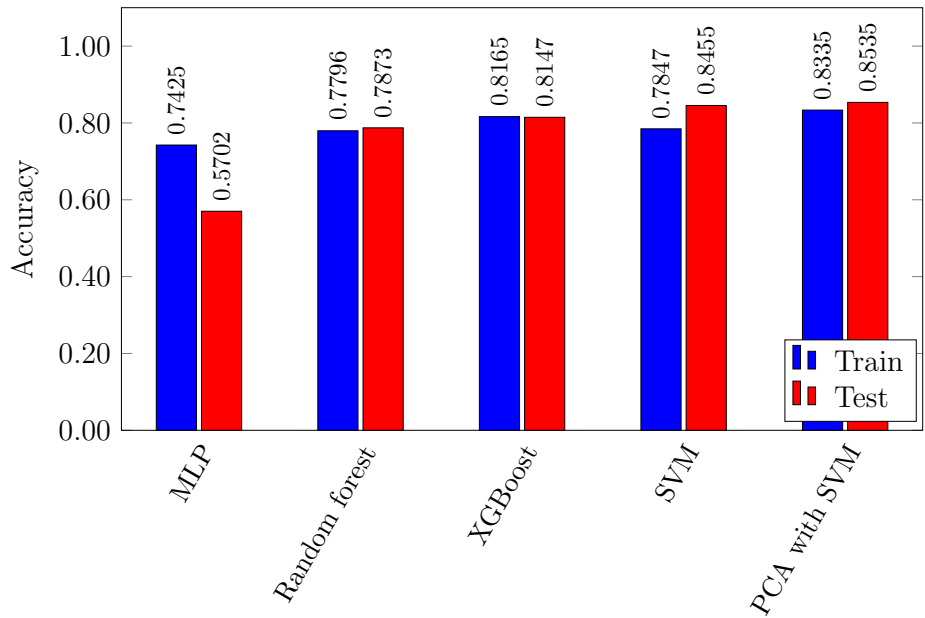

\centering
     \input figures/barBasic.tex
\caption{Accuracies for basic techniques}\label{fig:basic}
\end{figure}

Table~\ref{tab:basic} we provide a more detailed comparison of the
basic techniques considered above. The training and test accuracies
are also given in the form of a bar graph in Figure~\ref{fig:basic}.
We see that XGBoost, SVM, and PCA with SVM all yield
an test accuracy of~0.85. The high training accuracy for
random forest and XGBoost are indicative of overfitting.

\begin{table}[!htb]
\caption{Accuracy, precision, recall, and F1 score for basic techniques}
\label{tab:basic}
\centering
\adjustbox{scale=0.85}{
\begin{tabular}{cl|cccc}\midrule\midrule
Technique & Mode & Precision & Recall & F1 & Accuracy \\ \toprule
\multirow{2}{*}{MLP} & Train & 0.7425 & 1.0000 & 0.8523 & 0.7425 \\
				& Test &  0.7493 & 0.6326 & 0.6860 & 0.5702 \\ \midrule
\multirow{2}{*}{Random forest} & Train & 0.7800 & 0.9795 & 0.8684 & 0.7796 \\
				& Test & 0.7820 & 0.9893 & 0.8735 & 0.7873 \\ \midrule
\multirow{2}{*}{XGBoost} & Train & 0.8254 & 0.9548 & 0.8854 & 0.8165 \\
				& Test & 0.8171 & 0.9667 & 0.8856 & 0.8147 \\ \midrule
\multirow{2}{*}{SVM} & Train & 0.8510 & 0.8598 & 0.8554 & 0.7847 \\
				& Test & 0.8698 & 0.9301 & 0.8989 & 0.8455 \\ \midrule
\multirow{2}{*}{PCA with SVM} & Train & 0.9201 & 0.9675 & 0.9432 & 0.8335 \\
				& Test & 0.8567 & 0.9632 & 0.9068 & 0.8535 \\
				\midrule\midrule
\end{tabular}
}
\end{table}

\begin{figure}[!htb]
\centering
    \includegraphics[width=80mm]{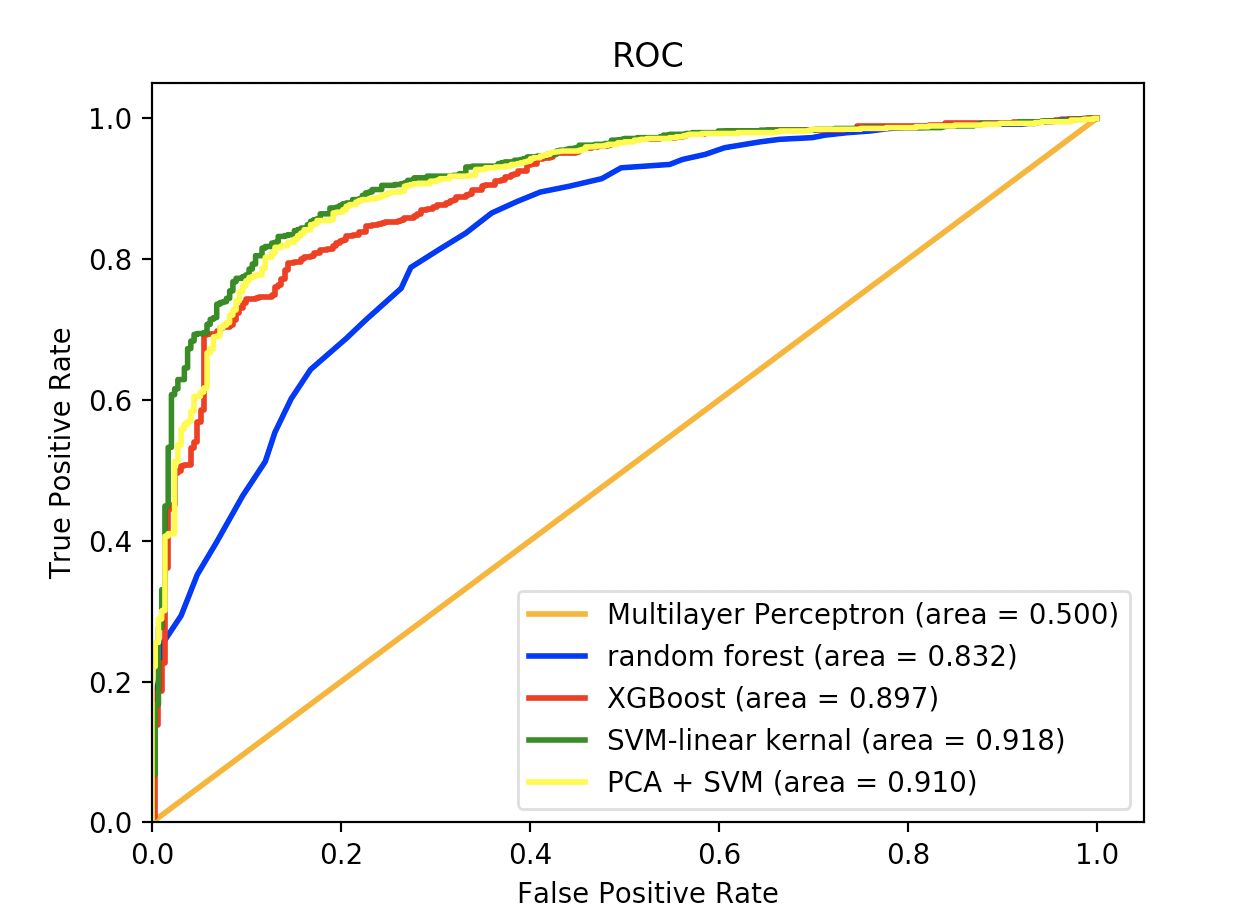}
\caption{ROC analysis of basic classifiers}\label{fig:ROC}
\end{figure}

Figure~\ref{fig:ROC} gives ROC curves for each of the basic classifiers
considered above. By the AUC measure, SVM performs best, marginally outperforming
the PCA with SVM model.

\subsection{Advanced Techniques}

In this section, we will discuss the performance of more advanced
learning techniques. Specifically, we consider CNN, AlexNet, DenseNet, ResNet, 
and a ResNet model that includes realtime data augmentation.

\subsubsection{Convolutional Neural Network}

\begin{figure}[!htb]
        \centering
        \includegraphics[width=80mm]{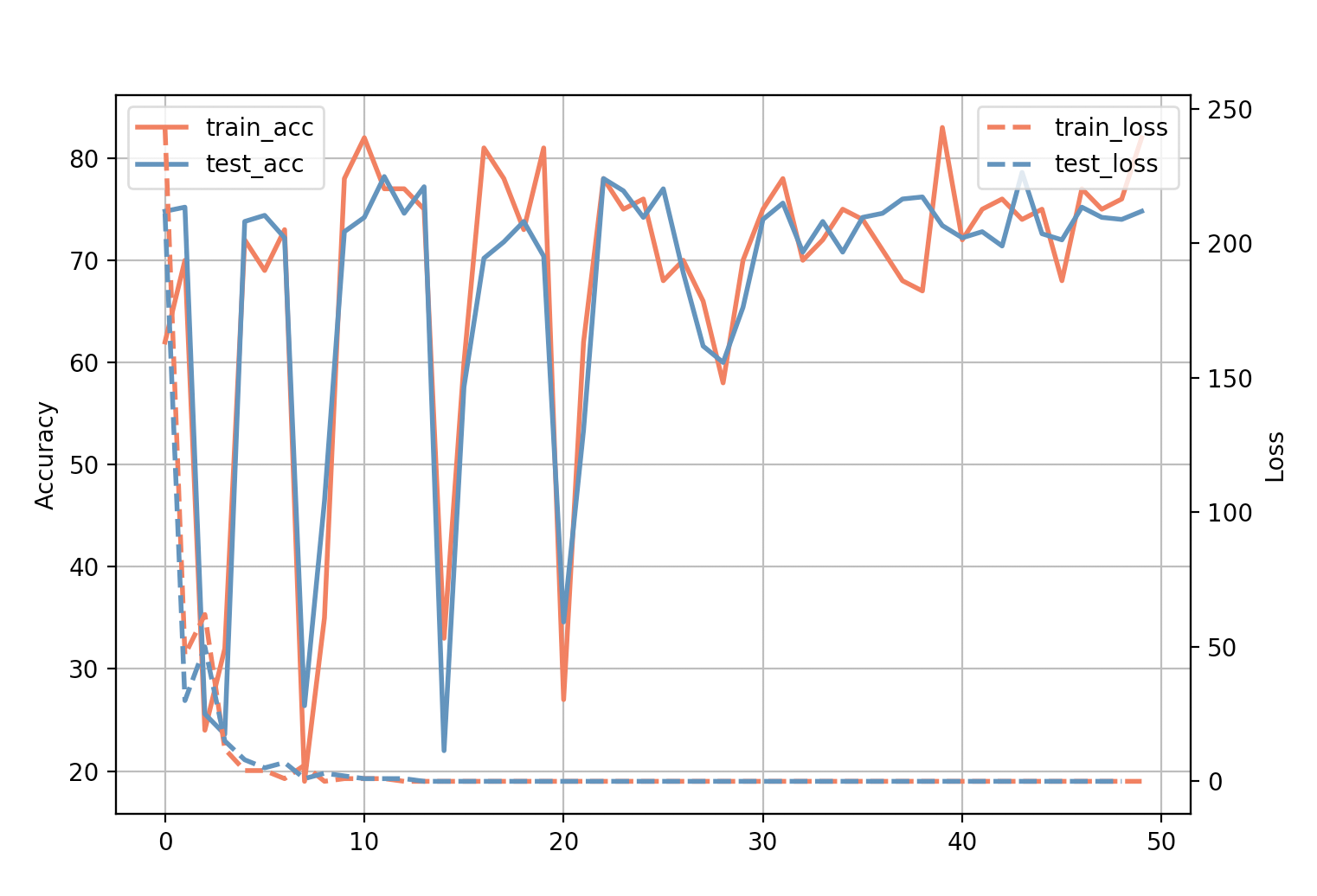}
        \caption{CNN performance}
        \label{fig:CNN result}
    \end{figure}

Our CNN model is implemented in \texttt{Tensorflow} 
and includes two convolution layers, a learning rate of~0.005, 
a \texttt{max\_pool} size of two, and a final fully connected layer. 
We experimented with various batch sizes, number of generation, 
and optimizers. The loss and accuracy plots for our 
best result (which used the Adam optimizer) 
are shown in Figure~\ref{fig:CNN result}.
In this case, we have a training accuracy of~0.8143
test accuracy of~0.7860.

\subsubsection{AlexNet}

\begin{figure}[!htb]
        \centering
        \includegraphics[width=80mm]{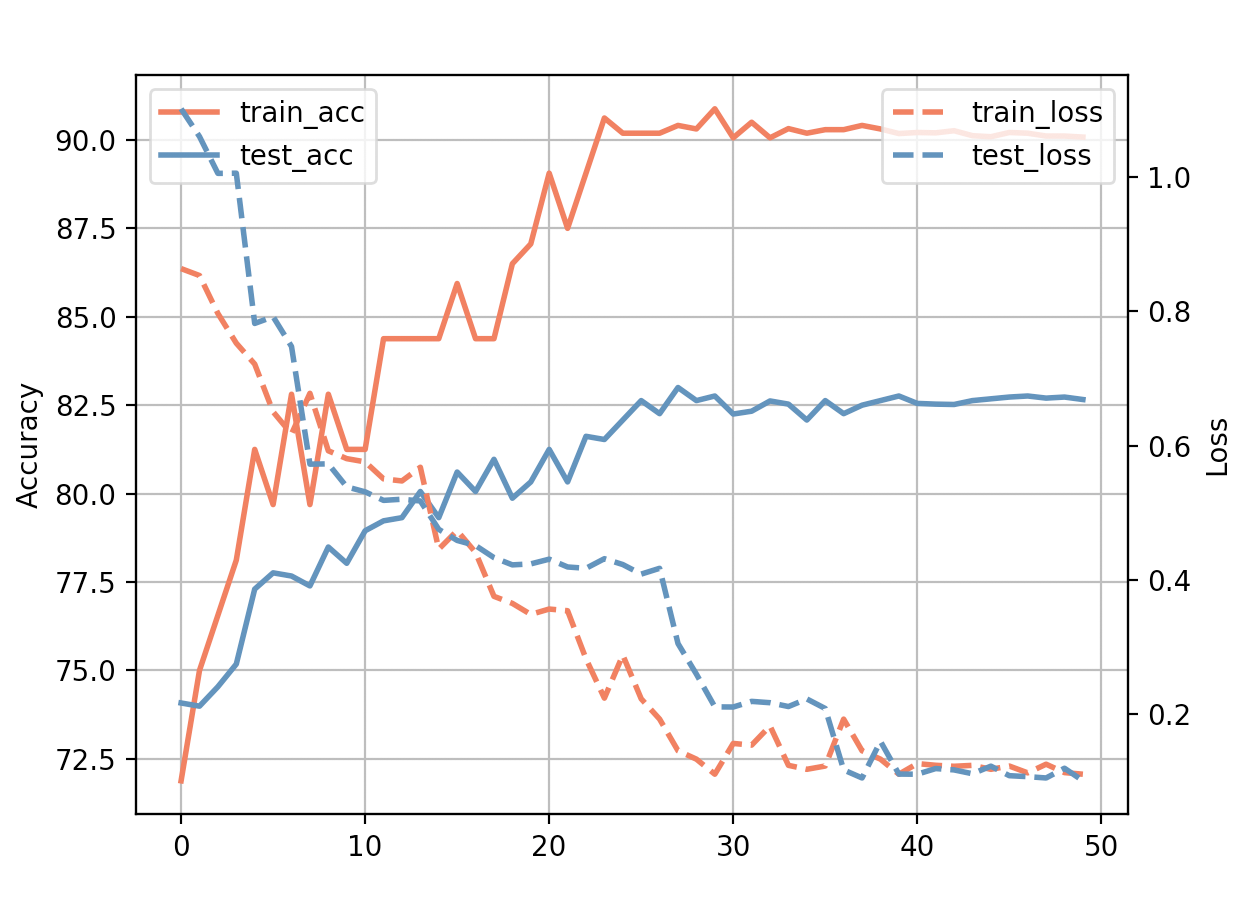}
        \caption{AlexNet performance}
        \label{fig:AlexNet Result}
    \end{figure}

We use the original AlexNet architecture with a learning rate of~$0.001$, 
softmax as the loss function, and the Adam optimizer. We trained the model 
for~50 epochs. The best train and test accuracy we attain with AlexNet
on our LGESS image dataset are ~0.9153 and ~0.8314. Figure~\ref{fig:AlexNet Result} shows 
the loss and accuracy plots for this case. 

\subsubsection{DenseNet}

\begin{figure}[!htb]
        \centering
        \includegraphics[width=80mm]{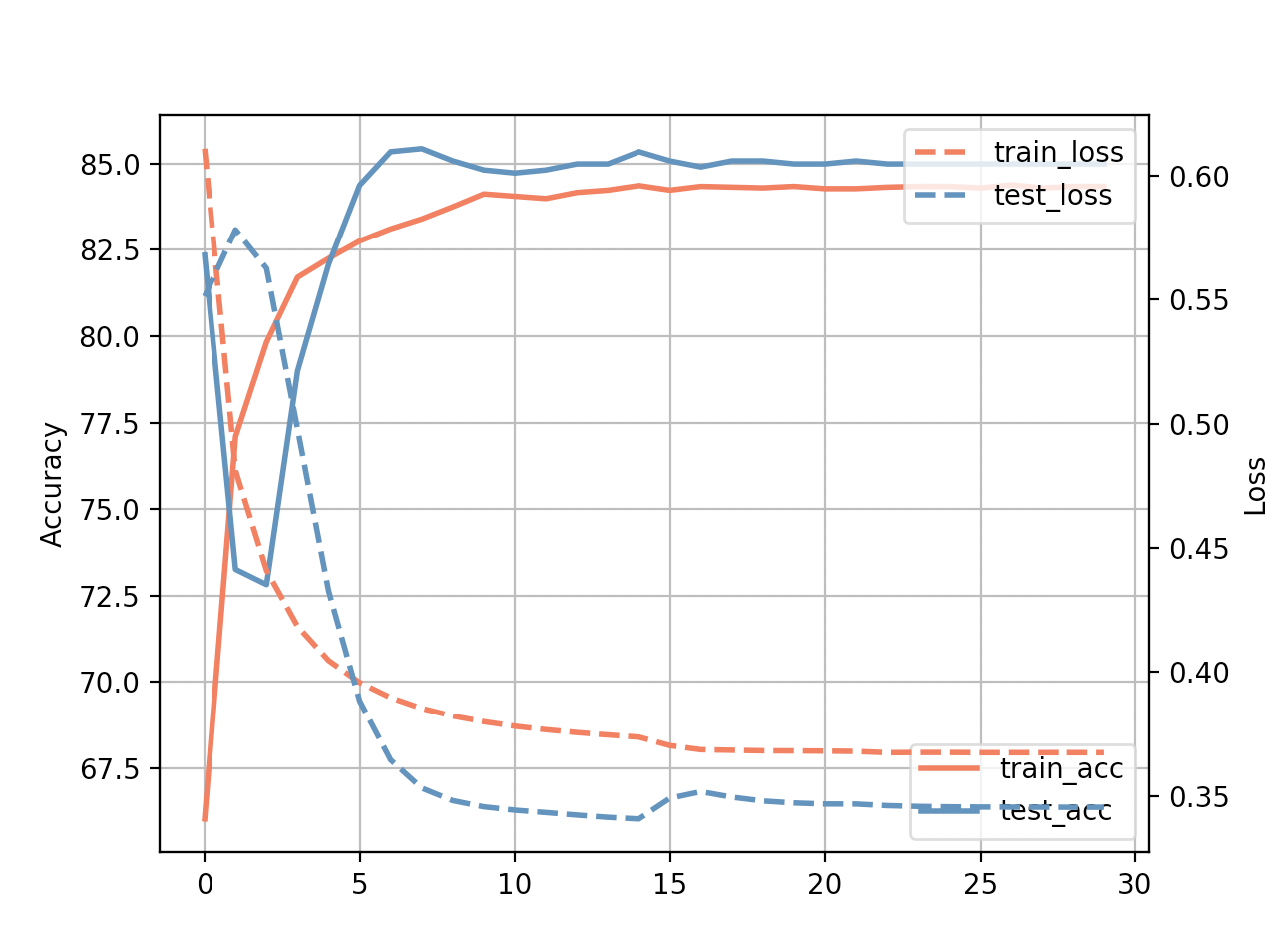}
        \caption{DenseNet performance}
        \label{fig:DenseNet result}
    \end{figure}

We built a DenseNet based on the \texttt{Keras} library. Each convolution layer is
followed by a max pooling layer. Three dense blocks are used,
and there is a classification layer at the end. We found that a learning rate to~0.001 
and~16 filters gave the best results. We trained the model for~30 epochs. 
This DenseNet, when applied to our LGESS dataset, yields a training accuracy of~0.8683 and a testing accuracy of~0.8457 Figure~\ref{fig:DenseNet result} details our DenseNet performance.

\subsubsection{ResNet}

\begin{figure}[!htb]
        \centering
        \includegraphics[width=80mm]{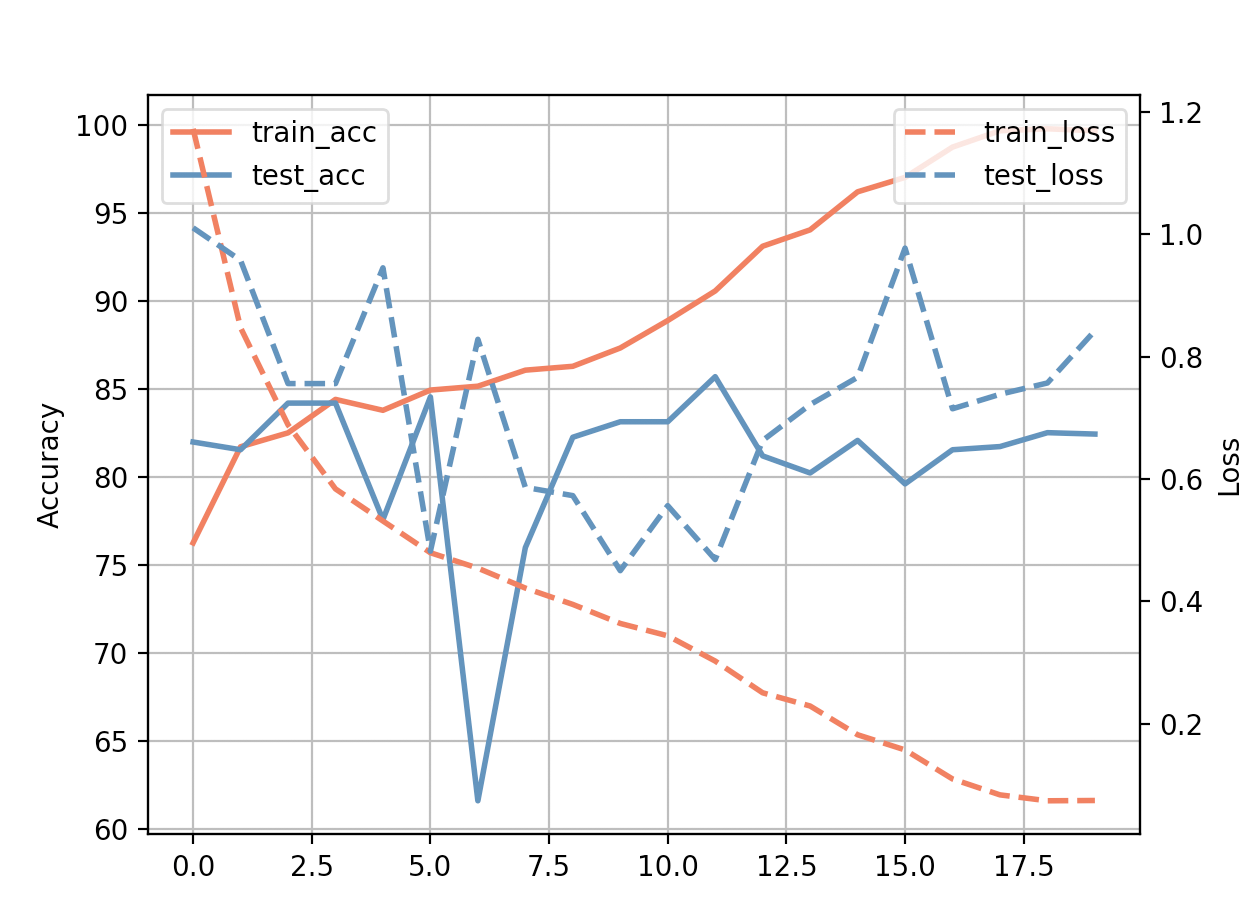}
        \caption{ResNet performance}
        \label{fig:ResNet result}
    \end{figure}

We also used the \texttt{Keras} library to build our ResNet. 
This ResNet is built with a convolution layer, max pooling layer, 
basic block layers, and an average pooling layer. The basic block layer 
includes a convolution layer, a batch normalization layer, and an activation layer. 
We trained our model for~20 epochs and obtained a best training accuracy 
of~0.9978, and testing accuracy of~0.8570. Figure~\ref{fig:ResNet result} shows the performance of 
this ResNet model on our dataset.

\subsubsection{ResNet with Realtime Data Augmentation}

\begin{figure}[!htb]
        \centering
        \includegraphics[width=80mm]{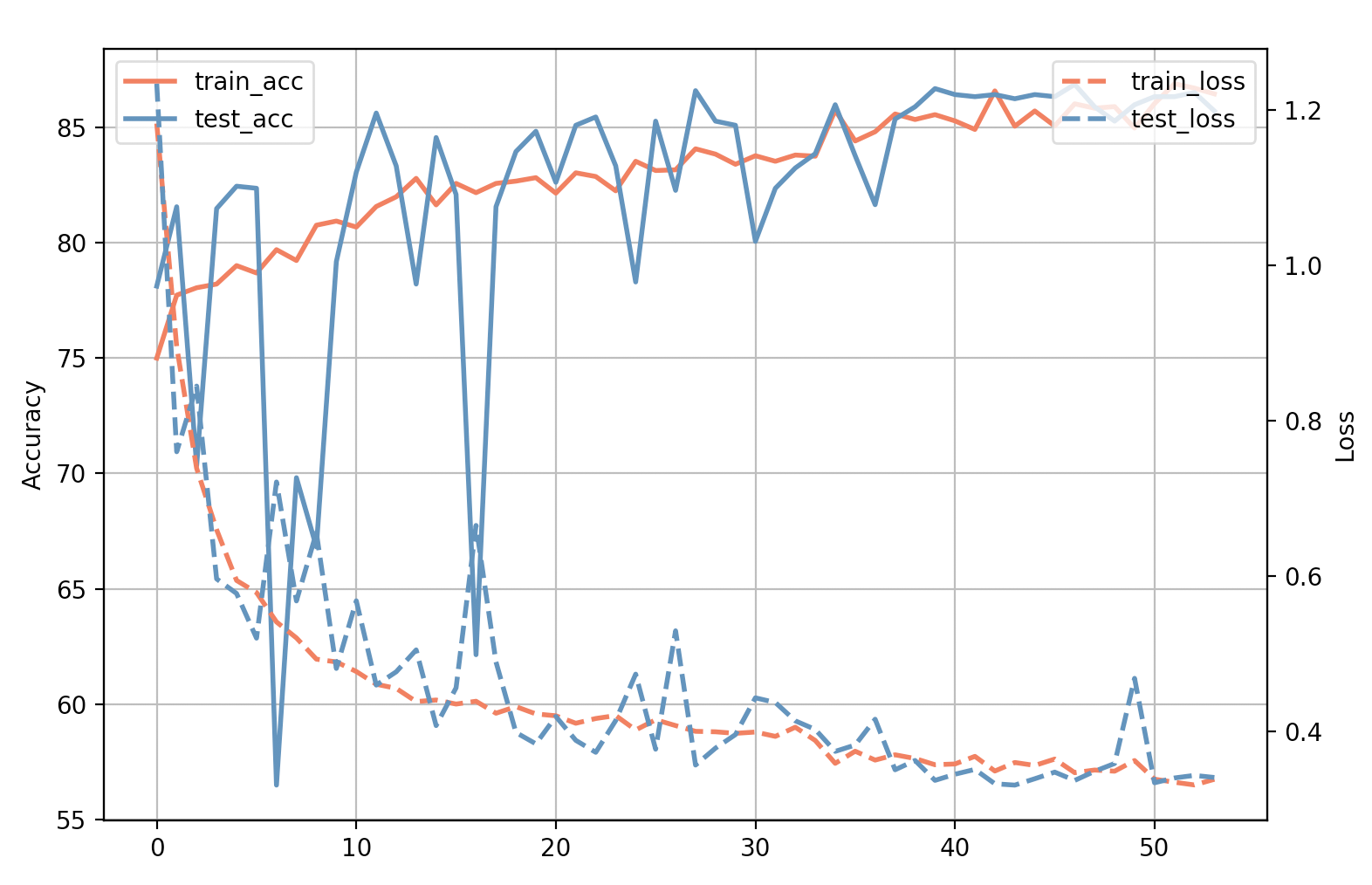}
        \caption{ResNet with data augmentation performance}
        \label{fig:ResNet with data}
    \end{figure}

Since ResNet showed the best performance, we decided to test 
ResNet with realtime data augmentation. In realtime data augmentation, 
images in the existing dataset are copied, randomly rotated, shifted, or flipped 
to form an augmented dataset. ResNet with data augmentation yielded a
slightly improved accuracy of~0.8788 in training and ~0.8685 in testing. 
The accuracy and loss plots for
this model are shown in Figure~\ref{fig:ResNet with data}.

\begin{figure}[!htb]
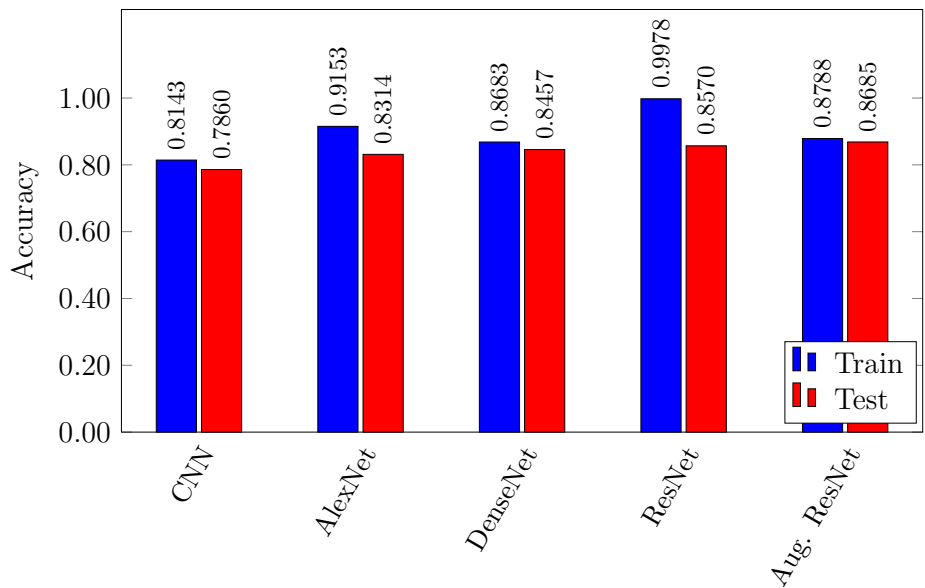

\centering
     \input figures/barAdvanced.tex
     \caption{Accuracies for advanced techniques}\label{fig:advanced}
\end{figure}

Figure~\ref{fig:advanced} compares the training and test accuracy 
for each advanced technique considered in this section. 
We note that ResNet with real time data augmentation has the highest classification 
accuracy at~0.8685. 

\subsection{Discussion}

\begin{figure}[!htb]
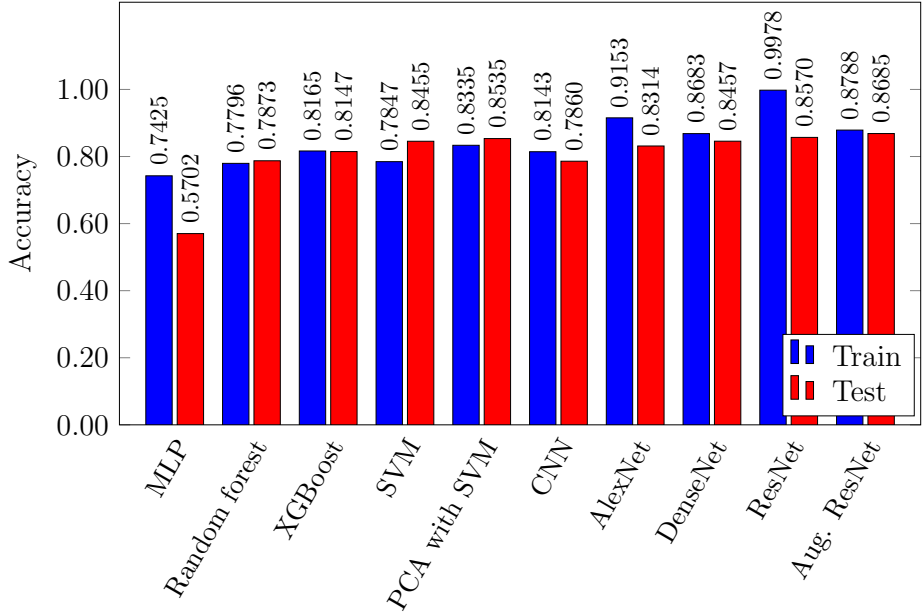

\centering
     \input figures/barAll.tex
     \caption{Test accuracy for all techniques}\label{fig:all}
\end{figure}

Figure~\ref{fig:all} summarizes the test performance of 
all of the machine learning and deep learning algorithm
we experimented with on our LGESS image datase. 
XGBoost, SVM and PCA with SVM all yield nearly identical
accuracy of~0.85, while SVM has a slight edge in terms of AUC.
ResNet yielded an accuracy of~0.8570, which was improved by about~0.01 
with realtime data augmentation. 

Comparing our results, we see that several models perform nearly-equally
well. Furthermore, accuracies in the range of~0.85 are much better than
is currently achieved in practice. 
 
\section{Conclusion}\label{chap:conclusion}

Cancer results in massive health care expenses and staggering 
death tolls~\cite{s}. Early detection of cancer can significantly reduce 
mortality. Therefore, fast, easy-to-use, and high precise tools 
for automatic cancer screening could lower healthcare costs, 
and ultimately save lives. 

We are not aware of previous research into the efficacy of machine learning 
and deep learning models in screening for LGESS. Our results indicate
that several models have the potential to provide accurate results in
a realistic setting. Specifically, we found that XGBoost, SVM, and ResNet 
models could achieve accuracies of~0.85 or better.

Today, 75\%\ of LGESS patients are incorrectly diagnosed with benign leiomyoma, 
leading to a delay in needed treatment and lower chances of survival. 
In this work, we have demonstrated that machine learning can,
at a minimum, serve as a second opinion, and thereby prevent 
many incorrect diagnoses.

In future studies, additional characteristics can be extracted from images, 
which would enable learning algorithms to work with a more refined and 
detailed feature space. Moreover, there exists a plethora of additional machine learning 
and deep learning algorithms that could be considered Such further research
could benefit individuals who suffer from the rare, but potentially deadly, disease
of low grade endometrial stromal sarcoma.

\bibliographystyle{plain}
\bibliography{references.bib}

\end{document}

%% file: preamble.tex
\usepackage{amsmath,amsthm, amsfonts, amssymb, amsxtra,amsopn}
\usepackage{pgfplots}
\usepgfplotslibrary{colormaps}
\pgfplotsset{compat=1.15}
\usepackage{pgfplotstable}
\usetikzlibrary{pgfplots.statistics} 
\usepackage{filecontents}
\usepackage{graphicx}
\usepackage{multirow}
\usepackage{tabu}
\usepackage{algorithmic}
\usepackage{longtable}
\usepackage{booktabs}
\usepackage{listings}
\usepackage{cmap}
\usepackage{colortbl}
\usepackage{adjustbox}
\usepackage{epsfig}
\usepackage{xstring}

\usepackage{subfigure}

\PassOptionsToPackage{hyphens}{url}

\usepackage{hyperref}
\hypersetup{colorlinks=true,linkcolor=black,citecolor=black,urlcolor=blue,filecolor=black}
\hypersetup{pdfpagemode=UseNone,pdfstartview=}

\usepackage[tableposition=top]{caption}
\usepackage{enumitem}
\setlist[itemize]{noitemsep, topsep=0pt}

\advance\oddsidemargin by -0.45in
\advance\textwidth by 0.9in

\advance\topmargin by -0.3in
\advance\textheight by 0.6in

\long\def\symbolfootnotetext[#1]#2{\begingroup%
\def\thefootnote{\fnsymbol{footnote}}\footnotetext[#1]{#2}\endgroup}

\pgfkeys{
    /pgf/number format/fixed zerofill=true }
    
\pgfplotstableset{
    color cells/.code={%
        \pgfqkeys{/color cells}{#1}%
        \pgfkeysalso{%
            postproc cell content/.code={%
                \begingroup
                \pgfkeysgetvalue{/pgfplots/table/@preprocessed cell content}\value
\ifx\value\empty
\endgroup
\else
                \pgfmathfloatparsenumber{\value}%
                \pgfmathfloattofixed{\pgfmathresult}%
                \let\value=\pgfmathresult
                \pgfplotscolormapaccess
                    [\pgfkeysvalueof{/color cells/min}:\pgfkeysvalueof{/color cells/max}]%
                    {\value}%
                    {\pgfkeysvalueof{/pgfplots/colormap name}}%
                \pgfkeysgetvalue{/pgfplots/table/@cell content}\typesetvalue
                \pgfkeysgetvalue{/color cells/textcolor}\textcolorvalue
                \toks0=\expandafter{\typesetvalue}%
                \xdef\temp{%
                    \noexpand\pgfkeysalso{%
                        @cell content={%
                            \noexpand\cellcolor[rgb]{\pgfmathresult}%
                            \noexpand\definecolor{mapped color}{rgb}{\pgfmathresult}%
                            \ifx\textcolorvalue\empty
                            \else
                                \noexpand\color{\textcolorvalue}%
                            \fi
                            \the\toks0 %
                        }%
                    }%
                }%
                \endgroup
                \temp
\fi
            }%
        }%
    }
}

\def\srowvecc#1#2{(\!\begin{array}{cc} 
      \noexpandarg\IfBeginWith{#1}{-}{\! #1}{#1}
    & #2\kern-0.5pt\end{array}\!)}
\def\rowvecc#1#2{\left(\!\begin{array}{cc} 
      \noexpandarg\IfBeginWith{#1}{-}{\! #1}{#1}
    & #2\kern-0.5pt\end{array}\!\right)}
\def\rowveccc#1#2#3{\left(\!\begin{array}{ccc} 
      \noexpandarg\IfBeginWith{#1}{-}{\! #1}{#1}
    & #2 
    & #3\kern-0.5pt\end{array}\!\right)}
\def\rowvecccc#1#2#3#4{\left(\!\begin{array}{cccc}
      \noexpandarg\IfBeginWith{#1}{-}{\! #1}{#1}
    & #2 
    & #3 
    & #4\kern-0.5pt\end{array}\!\right)}
\def\srowvecccc#1#2#3#4{\bigl(\!\begin{array}{cccc}
      \noexpandarg\IfBeginWith{#1}{-}{\! #1}{#1}
    & #2 
    & #3 
    & #4\kern-0.5pt\end{array}\!\bigr)}
\def\rowveccccc#1#2#3#4#5{\left(\!\begin{array}{ccccc} 
      \noexpandarg\IfBeginWith{#1}{-}{\! #1}{#1}
    & #2
    & #3
    & #4
    & #5\kern-0.5pt\end{array}\!\right)}
\def\srowvecccccc#1#2#3#4#5#6{(\!\begin{array}{cccccc} 
      \noexpandarg\IfBeginWith{#1}{-}{\! #1}{#1}
    & #2
    & #3
    & #4
    & #5
    & #6\kern-0.5pt\end{array}\!)}
\def\rowvecccccc#1#2#3#4#5#6{\left(\!\begin{array}{cccccc} 
      \noexpandarg\IfBeginWith{#1}{-}{\! #1}{#1}
    & #2
    & #3
    & #4
    & #5
    & #6\kern-0.5pt\end{array}\!\right)}

%% file: figures/conv3b.tex
\begin{tikzpicture}[scale=0.4]

\draw[red,ultra thick] (0.0,0.0) rectangle (11.2,11.2);

\draw[green,ultra thick] (0.5,-0.5) rectangle (11.7,10.7);

\draw[blue,ultra thick] (1.0,-1.0) rectangle (12.2,10.2);


\draw[ultra thick, dotted] (0.0,9.1) rectangle (2.1,11.2);
\draw[ultra thick] (0.0,11.2) -- (2.1,11.2);
\draw[ultra thick] (0.0,11.2) -- (0.0,9.1);
\draw[ultra thick] (1.0,8.1) rectangle (3.1,10.2);
\draw[ultra thick] (0.0,9.1) -- (1.0,8.1);
\draw[ultra thick] (2.1,11.2) -- (3.1,10.2);
\draw[ultra thick] (0.0,11.2) -- (1.0,10.2);
\draw[ultra thick, dotted] (3.1,8.1) -- (2.1,9.1);

\draw[black,ultra thick] (13.6,1.4) rectangle (23.4,11.2);
\draw[ultra thick] (13.6,11.2) rectangle (14.3,10.5);
\draw[black,ultra thick] (14.1,0.9) rectangle (23.9,10.7);
\draw[ultra thick] (14.1,10.7) rectangle (14.8,10.0);
\draw[black,ultra thick] (14.6,0.4) rectangle (24.4,10.2);
\draw[ultra thick] (14.6,10.2) rectangle (15.3,9.5);
\draw[black,ultra thick] (15.1,-0.1) rectangle (24.9,9.7);
\draw[ultra thick] (15.1,9.7) rectangle (15.8,9.0);
\draw[black,ultra thick] (15.6,-0.6) rectangle (25.4,9.2);
\draw[ultra thick] (15.6,9.2) rectangle (16.3,8.5);

\draw[smooth,rounded corners,thick,->] (2.5,10.7) -- (2.5,11.9) -- (3.0,12.2) -- (12.95,12.2)  
	-- (13.55,11.9) -- (13.95,11.2);
\node at (7,12.7) {$\mbox{\footnotesize filter}_1$};
\draw[smooth,rounded corners,thick,->] (2.0,10.7) -- (2.0,12.9) -- (2.5,13.2) -- (12.95,13.2)  
	-- (13.55,12.9) -- (14.45,10.7);
\node at (7,13.7) {$\mbox{\footnotesize filter}_2$};
\draw[smooth,rounded corners,thick,->] (1.5,10.7) -- (1.5,13.9) -- (2.0,14.2) -- (12.95,14.2)  
	-- (13.55,13.9) -- (14.95,10.2);
\node at (7,14.7) {$\mbox{\footnotesize filter}_3$};
\draw[smooth,rounded corners,thick,->] (1.0,10.7) -- (1.0,14.9) -- (1.5,15.2) -- (12.95,15.2)  
	-- (13.55,14.9) -- (15.45,9.7);
\node at (7,15.7) {$\mbox{\footnotesize filter}_4$};
\draw[smooth,rounded corners,thick,->] (0.5,10.7) -- (0.5,15.9) -- (1.0,16.2) -- (12.95,16.2)  
	-- (13.55,15.9) -- (15.95,9.2);
\node at (7,16.7) {$\mbox{\footnotesize filter}_5$};

\end{tikzpicture}

%% file: figures/barBasic.tex
\begin{tikzpicture}[scale=0.95, every node/.style={scale=1.0}]
\begin{axis}[
        width  = 0.8*\textwidth,
        height = 7.5cm,
        ymin=0,ymax=1.10,
        ytick={0.00,0.20,0.40,0.60,0.80,1.00},
        major x tick style = transparent,
        ybar=5*\pgflinewidth,
        bar width=16.0pt,
        ylabel = {Accuracy},
        symbolic x coords={MLP, Random forest, XGBoost, SVM, PCA with SVM},
        xticklabels={MLP, Random forest, XGBoost, SVM, PCA with SVM},
	y tick label style={
    		/pgf/number format/.cd,
   		fixed,
   		fixed zerofill,
    		precision=2},
        xtick = data,
        x tick label style={
        		rotate=60,
		font=\small,
		anchor=north east,
		inner sep=0mm
		},
        nodes near coords,
        every node near coord/.append style={rotate=90, 
        								   anchor=west,
								   font=\footnotesize,
								   /pgf/number format/.cd,
								   fixed,
								   fixed zerofill,
								   precision=4},
        enlarge x limits=0.12,
        legend cell align=left,
        legend style={
                at={(0.91,0.02)},
                anchor=south,
                column sep=1ex
        },
]
\addplot [fill=blue,opacity=1.00]
coordinates {
(MLP, 0.7425)
(Random forest, 0.7796)
(XGBoost, 0.8165)
(SVM, 0.7847)
(PCA with SVM, 0.8335)
};
\addlegendentry{Train}
\addplot [fill=red,opacity=1.00]
coordinates {
(MLP, 0.5702)
(Random forest, 0.7873)
(XGBoost, 0.8147)
(SVM, 0.8455)
(PCA with SVM, 0.8535)
};
\addlegendentry{Test}
\end{axis}
\end{tikzpicture}

%% file: figures/barAdvanced.tex
\begin{tikzpicture}[scale=0.95, every node/.style={scale=1.0}]
\begin{axis}[
        width  = 0.8*\textwidth,
        height = 7.5cm,
        ymin=0,ymax=1.265,
        ytick={0.00,0.20,0.40,0.60,0.80,1.00},
        major x tick style = transparent,
        ybar=5*\pgflinewidth,
        bar width=16.0pt,
        ylabel = {Accuracy},
        symbolic x coords={CNN, AlexNet, DenseNet, ResNet, ResNet with augmentation},
        xticklabels={CNN, AlexNet, DenseNet, ResNet, Aug. ResNet},
	y tick label style={
    		/pgf/number format/.cd,
   		fixed,
   		fixed zerofill,
    		precision=2},
        xtick = data,
        x tick label style={
        		rotate=60,
		font=\small,
		anchor=north east,
		inner sep=0mm
		},
        nodes near coords,
        every node near coord/.append style={rotate=90, 
        								   anchor=west,
								   font=\footnotesize,
								   /pgf/number format/.cd,
								   fixed,
								   fixed zerofill,
								   precision=4},
        enlarge x limits=0.12,
        legend cell align=left,
        legend style={
                at={(0.91,0.02)},
                anchor=south,
                column sep=1ex
        },
]
\addplot [fill=blue,opacity=1.00]
coordinates {
(CNN, 0.8143)
(AlexNet, 0.9153)
(DenseNet, 0.8683)
(ResNet, 0.9978)
(ResNet with augmentation, 0.8788)
};
\addlegendentry{Train}
\addplot [fill=red,opacity=1.00]
coordinates {
(CNN, 0.7860)
(AlexNet, 0.8314)
(DenseNet, 0.8457)
(ResNet, 0.8570)
(ResNet with augmentation, 0.8685)
};
\addlegendentry{Test}
\end{axis}
\end{tikzpicture}

%% file: figures/barAll.tex
\begin{tikzpicture}[scale=0.95, every node/.style={scale=1.0}]
\begin{axis}[
        width  = 0.8*\textwidth,
        height = 7.5cm,
        ymin=0,ymax=1.26,
        ytick={0.00,0.20,0.40,0.60,0.80,1.00},
        major x tick style = transparent,
        ybar=5*\pgflinewidth,
        bar width=10.5pt,
        ylabel = {Accuracy},
        symbolic x coords={MLP, Random forest, XGBoost, SVM, PCA with SVM,
        CNN, AlexNet, DenseNet, ResNet, ResNet with augmentation},
        xticklabels={MLP, Random forest, XGBoost, SVM, PCA with SVM,
        CNN, AlexNet, DenseNet, ResNet, Aug. ResNet},
	y tick label style={
    		/pgf/number format/.cd,
   		fixed,
   		fixed zerofill,
    		precision=2},
        xtick = data,
        x tick label style={
        		rotate=60,
		font=\small,
		anchor=north east,
		inner sep=0mm
		},
        nodes near coords,
        every node near coord/.append style={rotate=90, 
        								   anchor=west,
								   font=\footnotesize,
								   /pgf/number format/.cd,
								   fixed,
								   fixed zerofill,
								   precision=4},
        enlarge x limits=0.08,
        legend cell align=left,
        legend style={
                at={(0.91,0.02)},
                anchor=south,
                column sep=1ex
        },
]
\addplot [fill=blue,opacity=1.00]
coordinates {
(MLP, 0.7425)
(Random forest, 0.7796)
(XGBoost, 0.8165)
(SVM, 0.7847)
(PCA with SVM, 0.8335)
(CNN, 0.8143)
(AlexNet, 0.9153)
(DenseNet, 0.8683)
(ResNet, 0.9978)
(ResNet with augmentation, 0.8788)
};
\addlegendentry{Train}
\addplot [fill=red,opacity=1.00]
coordinates {
(MLP, 0.5702)
(Random forest, 0.7873)
(XGBoost, 0.8147)
(SVM, 0.8455)
(PCA with SVM, 0.8535)
(CNN, 0.7860)
(AlexNet, 0.8314)
(DenseNet, 0.8457)
(ResNet, 0.8570)
(ResNet with augmentation, 0.8685)
};
\addlegendentry{Test}
\end{axis}
\end{tikzpicture}